\documentclass[english,reprint, longbibliography, superscriptaddress, breaklinks=true, showkeys, showpacs=false, nofootinbib]{revtex4}
\usepackage[T1]{fontenc}
\usepackage[latin9]{inputenc}
\usepackage{color}
\usepackage{babel}
\usepackage{amsmath}
\usepackage{amsthm}
\usepackage{amsfonts}
\usepackage{amssymb}
\usepackage{graphicx}
\usepackage{subfigure}
\usepackage{physics}

\usepackage[unicode=true,pdfusetitle,
 bookmarks=true,bookmarksnumbered=false,bookmarksopen=false,
 breaklinks=true,pdfborder={0 0 0},backref=false,colorlinks=true]
 {hyperref} 
\usepackage{breakurl}

\makeatletter
\@ifundefined{textcolor}{}
{%
 \definecolor{BLACK}{gray}{0}
 \definecolor{WHITE}{gray}{1}
 \definecolor{RED}{rgb}{1,0,0}
 \definecolor{GREEN}{rgb}{0,1,0}
 \definecolor{BLUE}{rgb}{0,0,1}
 \definecolor{CYAN}{cmyk}{1,0,0,0}
 \definecolor{MAGENTA}{cmyk}{0,1,0,0}
 \definecolor{YELLOW}{cmyk}{0,0,1,0}
}

\pdfoutput=1
\hypersetup{colorlinks=true,citecolor=blue,linkcolor=cyan,urlcolor=blue,filecolor= green, breaklinks=true}
\usepackage{url}
\usepackage{breakurl}

\makeatother

\begin{document}

\title{Complete complementarity relations and its Lorentz invariance}

\author{Marcos L. W. Basso}
\email{marcoslwbasso@mail.ufsm.br}
\address{Departamento de F\'isica, Centro de Ci\^encias Naturais e Exatas, Universidade Federal de Santa Maria, Avenida Roraima 1000, Santa Maria, Rio Grande do Sul, 97105-900, Brazil}

\author{Jonas Maziero}
\email{jonas.maziero@ufsm.br}
\address{Departamento de F\'isica, Centro de Ci\^encias Naturais e Exatas, Universidade Federal de Santa Maria, Avenida Roraima 1000, Santa Maria, Rio Grande do Sul, 97105-900, Brazil}

\selectlanguage{english}%

\begin{abstract}
It is well known that entanglement under Lorentz boosts is highly dependent on the boost scenario in question. For single particle states, a spin-momentum product state can be transformed into an entangled state. However, entanglement is just one of the aspects that completely characterizes a quantum system. The other two are known as the wave-particle duality. Although the entanglement entropy does not remain invariant under Lorentz boosts, and neither do the measures of predictability and coherence, we show here that these three measures taken together, in a complete complementarity relation (CCR), are Lorentz invariant. Peres et al., in [Phys. Rev. Lett. 88, 230402 (2002)], realized that even though it is possible to formally define spin in any Lorentz frame, there is no relationship between the observable expectation values in different Lorentz frames.  Analogously, one can, in principle, define complementary relations in any Lorentz frame, but there is no obvious transformation law relating complementary relations in different frames. However, our result shows that the CCR have the same value in any Lorentz frame, i.e., there's a transformation law connecting the complete complementarity relations. In addition, we explore relativistic scenarios for single and two particle states, which helps in understanding the exchange of different aspects of a quantum system under Lorentz boosts.
\end{abstract}


\maketitle

\section{Introduction}
Entanglement is one of the most intriguing characteristics that turns apart the quantum world from the classical world. Its fundamental importance in quantum foundations \cite{Schrodinger, Hor}, together with its application in several areas, such as quantum information and quantum computation \cite{Popescu, Preskill, Cavalcanti}, has made the entanglement theory achieve great progress in recent decades. Moreover, there has been more and more interest in how entanglement behaves under relativistic settings \cite{Terno}. For instance, in Ref. \cite{Czachor} the author considered the relativistic version of the famous Einstein-Podolsky-Rosen experiment with massive spin-1/2 particles. Czachor argued that the degree of violation of the Bell inequality is dependent on the velocity of the particles, leading to implications for quantum cryptography. A few years later, the authors in Refs. \cite{Adami, Ueda} showed that the entanglement of Bell states depends on the velocity of an observer. On the other hand, the authors of  Ref. \cite{Milburn} argued that the entanglement fidelity of a Bell state remains invariant for a Lorentz boosted observer. However, in the same year, it was demonstrated by Peres et al. \cite{Peres} that the entropy of a single massive spin-1/2 particle does not remain invariant under Lorentz boosts. Thereafter, the behavior of entanglement under Lorentz boosts has been receiving a lot of attention by researchers \cite{Bergo, Li, Moon, Lee, Jordan, Vlatko, Friis, Nasr, Blasone}.

As pointed out by Palge and Dunningham in Ref. \cite{Palge}, the main aspect to be noticed here is that many of these apparently conflicting results involve systems containing different particle states and boost geometries. Therefore, entanglement under Lorentz boosts is highly dependent on the boost scenario in question \cite{Dunningham}. For single particle states, a spin-momentum product state can be transformed into an entangled state. Beyond that, Lorentz boosts can be regarded as controlled quantum operations where momentum plays the role of the control system, whereas the spin can be taken as the target qubit, as argued in Ref. \cite{Vlatko}. This implies that Lorentz boosts perform global transformations on single particle systems. As in Refs. \cite{Jordan, Friis, Palge}, by using discrete momentum states, in this article we discuss the fact that for a spin-momentum product state be transformed into a entangled state it needs coherence between the momentum states. Otherwise, if the momentum state is completely predictable, the spin-momentum state remains separable, and the Lorentz boost will at most generate superposition between the spin states.
In addition, we discuss similar results for the two-particle states under Lorentz boosts. As already noticed in Refs. \cite{Adami, Palge}, the state and entanglement changes of the different degrees of freedom depend considerably on the initial states involved, as well as on the geometry of the boost scenario. Whereas some states and geometries leave the overall entanglement invariant, others create entanglement.

Besides, it is known that entanglement is just one of the aspects that completely characterizes a quanton \cite{Leblond}. The other two, which also are intriguing characteristics that turn apart the quantum world from the classical world, are known as the wave-particle duality. This distinguished aspect is generally captured, in a qualitative way, by Bohr's complementarity principle \cite{Bohr}. For instance, in the Mach-Zehnder interferometer or in the double-slit interferometer,  the wave aspect is characterized by interference fringes visibility, meanwhile the particle nature is given by the which-way information of the path along the interferometer. A quantitative version of the wave-particle duality was first investigated by Wooters and Zurek \cite{Wootters}, and later captured by a complementarity inequality in Refs. \cite{Engle, Yasin}:
\begin{equation}
    P^2 + V^2 \le 1, \label{eq:cr1}
\end{equation}
where $P$ is an \textit{a priori} predictability, for which the particle aspect is inferred once the quanton is more likely to follow one path than the other and it is directly related to the probability distribution given by the diagonal elements of the density operator as we will discuss in further section. Besides, $V$ is the visibility of the interference pattern. Recently, several steps have been taken towards the quantification of the wave-particle duality, with the establishment of minimal and reasonable conditions that visibility and predictability measures should satisfy \cite{Durr, Englert}. As well, with the development of the field of quantum information, it was suggested that the quantum coherence \cite{Baumgratz} would be a good generalization of the visibility measure \cite{Bera, Bagan, Tabish, Mishra}. Until now, many approaches were taken for quantifying the wave-particle properties of a quantum system \cite{Angelo, Coles, Hillery, Qureshi, Maziero}. As pointed out by Qian et al. \cite{Qian}, complementarity relations like Eq. (\ref{eq:cr1}) do not really predict a balanced exchange between $P$ and $V$ simply because the inequality permits a decrease of $P$ and $V$ together, or an increase by both. It even allows the extreme case $P = V = 0$ to occur (neither wave or particle) while, in an experimental setup, we still have a quanton on hands. Such a quanton cannot be nothing. Thus, one can see that something must be missing from Eq. (\ref{eq:cr1}). As noticed by Jakob and Bergou \cite{Janos}, this lack of knowledge about the system is due to entanglement, or, more generally, to quantum correlations \cite{Marcos}. This means that the information is being shared with another system and this kind of quantum correlation can be seen as responsible for the loss of purity of each subsystem such that, for pure maximally entangled states, it is not possible to obtain information about the local properties of the subsystems. Hence, to completely quantify a quanton, one has also to regard its correlations with other systems, such that the entire system is pure. 

In this paper, we study how these different aspects of a quanton behave under Lorentz boosts. Even though entanglement entropy does not remain invariant under Lorentz boosts, and neither do measures of predictability and coherence, we show that these three measures together, in what is known as a complete complementarity relation (CCR), are Lorentz invariant. In Ref. \cite{Peres}, the authors showed that, even though it is possible to formally define spin in any Lorentz frame, there is no relationship between the observable expectation values in different Lorentz frames. Here the situation is different. First, one can define complementary relations in any Lorentz frame, but there is no obvious transformation law relating complementary relations in different frames. However since the purity of a state is preserved under transformations between inertial frames, the complementary relations have the same value in any Lorentz frame, i.e., there is a transformation law connecting the complete complementarity relations. In addition, we explore several relativistic scenarios for single and two particle states, what helps in understanding the exchange of these different aspects of a quanton under Lorentz boosts. 

The organization of this article is as follows. In Sec. \ref{sec:rep}, we discuss the representations of the Poincar\'e group in the Hilbert space, as well as the Wigner's little group, by focusing in spin-$1/2$ massive particles. In Sec. \ref{sec:lorccr}, we obtain complete complementarity relations for multipartite pure quantum systems, and show that CCR are Lorentz invariant. Thereafter, in Sec. \ref{sec:relset}, we turn to
the study of the behavior of CCR in relativistic scenarios for several single and two particle states. Lastly, in Sec. \ref{sec:con}, we give our conclusions.

\section{Representations of the Poincar\'e group in the Hilbert space}
\label{sec:rep}
One of the fundamental questions when studying the relativistic formulation of the quantum theory is how quantum states behave under Lorentz boosts. In the language of group theory, we are seeking to represent an element of the Lorentz group by a unitary operator on the Hilbert space that the quantum states belongs to. More specifically, single particle quantum states are classified by their transformation under the inhomogeneous Lorentz group, or Poincar\'e group, which consists of homogeneous Lorentz transformations $\Lambda$ and translations $a$ \cite{Weinberg}. For our discussion, we adopt the following notation: Greek indices run over the $4$-spacetime coordinate labels $\{0, 1, 2, 3\}$; Latin indices run over the three spacial coordinates labels $\{1, 2, 3\}$; the Minkowski metric $\eta_{\mu \nu}$ is diagonal with elements $\{-1, 1, 1, 1\}$; $4$-vectors are in un-boldfaced type while spacial vectors are represented by an arrow. For instance, the $4$-momentum for a particle with mass $m$ is given by $p = (p^0, p^1, p^2, p^3) = (p^0, \vec{p})$, with norm $p^2 := p_{\mu}p^{\mu} = \eta_{\mu, \nu}p^{\nu}p^{\mu} = -(p^0)^2 + \vec{p}^2 = - m^2$, where we use natural units, i.e., $c = \hbar = 1$.

An inertial reference frame $\mathcal{O}$ is related to another inertial frame $\mathcal{O}'$ via a Poincar\'e transformation
\begin{equation}
    x'^{\mu} := T(\Lambda,a)x^{\nu} = \Lambda^{\mu}_{\nu} x^{\nu} + a^{\mu},
\end{equation}
with $x = (x^0, \vec{x})$ being the coordinates of $\mathcal{O}$, and similarly for $\mathcal{O}'$. Then $T(\Lambda,a)$ induces a unitary transformation on quantum states characterized by
\begin{equation}
    \ket{\Psi} \to U(\Lambda,a) \ket{\Psi},
\end{equation}
which satisfies the same composition rule of $T(\Lambda,a)$:
\begin{equation}
    U(\Lambda_1,a_1)U(\Lambda_2,a_1) = U(\Lambda_1 \Lambda_2, \Lambda_1 a_2 + a_1). 
\end{equation}
Single particle quantum states can be denoted by $\ket{p} \otimes \ket{\sigma} := \ket{p, \sigma}$, where $p$ labels the $4$-momenta and $\sigma$ labels the spin for massive particles. The quantum states $\ket{p, \sigma}$ are eigenvectors of the momentum operator $P^{\mu}$ with eigenvalues $p^{\mu}$, i.e., $P^\mu \ket{p,\sigma} = p^{\mu} \ket{p, \sigma}$. This corresponds to a basis of plane waves and, thus, transforms under translations as $U(I,a) \ket{p, \sigma} := U(a)\ket{p, \sigma} = e^{-ipa} \ket{p, \sigma}$, where $pa := p_{\mu}a^{\mu} = p^{\mu}a_{\mu}$. Meanwhile, a general Lorentz boost $\Lambda$ takes the eigenvalue $p^{\mu} \to \Lambda^{\mu}_{\nu} p^{\nu}$, and therefore $U(\Lambda,0)\ket{p, \sigma} := U(\Lambda)\ket{p, \sigma}$ must be a linear combination of all states with momentum $\Lambda p$, i.e.,
\begin{equation}
    U(\Lambda)\ket{p, \sigma} = \sum_{\lambda} D_{\lambda, \sigma}(\Lambda, p) \ket{\Lambda p, \lambda}.
\end{equation}
As $U(\Lambda)$ is a representation, it preserves the group structure, and  imposes
conditions on the values of $D_{\lambda, \sigma}$. To see this, let us recall that $ U(\Lambda)$ leaves $p^2 := p_{\mu}p^{\mu} = \vec{p} \cdot \vec{p} - E^2= -m^2$ and the sign of $p^0 = E$ unchanged for a particle with mass $m$. Hence, we can use these two invariant quantities to classify states into specific classes. For each value of $p^2$ and for each $sign(p^0)$, it is possible to choose a `standard' $4$-momentum $k$ that identifies a specific class of quantum states \cite{Lanzagorta}. For massive particles, we can fix the standard momentum $k$ to be the particle's momentum in the rest frame, i.e., $k = (m,0,0,0)$. Then, any momenta $p$ can be expressed in terms of the standard momentum, i.e., $p^{\mu} = (L(p)k)^{\mu} = L(p)^{\mu}_{\nu}k^{\nu}$, where $L(p)$ is a Lorentz transformation which depends on $p$ and takes $k \to p$. Therefore, quantum states $\ket{p, \sigma}$ can be defined in terms of the standard momentum state $\ket{k, \sigma}$:
\begin{equation}
    \ket{p, \sigma} = U(L(p)) \ket{k, \sigma}.
\end{equation}
Now, if we apply a Lorentz boost $\Lambda$ on $\ket{p, \sigma}$, then
\begin{align}
    U(\Lambda) \ket{p, \sigma} & = U(\Lambda)U(L(p)) \ket{k,\sigma} \\
    & = U(I)U(\Lambda L(p)) \ket{k, \sigma}\\
    & = U(L(\Lambda p) L^{-1}(\Lambda p))U(\Lambda L(p)) \ket{k, \sigma}\\
    & = U(L(\Lambda p)) U(L^{-1}(\Lambda p)\Lambda L(p)) \ket{k, \sigma} \\
    & = U(L(\Lambda p)) U(W(\Lambda,p))\ket{k,\sigma},
\end{align}
where $W(\Lambda,p) = L^{-1}(\Lambda p)\Lambda L(p)$ is called Wigner rotation, which leaves the standard momentum $k$ invariant, and  only acts on the internal degrees of freedom of $\ket{k, \sigma}$: $ k \xrightarrow{L} p \xrightarrow{\Lambda} \Lambda p \xrightarrow{L^{-1}} k$. Hence, the final momentum in the rest frame is different from the original one by a Wigner rotation, i.e., $U(W(\Lambda,p))\ket{k,\sigma} = \sum_{\lambda} D_{\lambda, \sigma} (W(\Lambda, p)) \ket{k, \lambda}$. On the other hand, $U(L(\Lambda p))$ takes $k \to \Lambda p$ without affecting the spin, by definition. Therefore,
\begin{align}
    U(\Lambda) \ket{p, \sigma} & =  U(L(\Lambda p)) U(W(\Lambda,p))\ket{k,\sigma} \\
    & = U(L(\Lambda p)) \sum_{\lambda} D_{\lambda, \sigma}W(\Lambda, p)) \ket{k, \lambda} \\
    & = \sum_{\lambda} D_{\lambda,\sigma} (W(\Lambda, p)) \ket{\Lambda p, \lambda}.
\end{align}
It is worth mentioning that the subscripts of $D_{\lambda,\sigma} (W(\Lambda, p))$ can be suppressed, and we can write $U(\Lambda) \ket{p, \sigma} = \ket{\Lambda p} \otimes D (W(\Lambda, p)) \ket{\sigma}$. The set of Wigner rotations forms a group known as the \textit{little group}, which is a subgroup of the Poincar\'e group \cite{Onuki}. In other words, under a Lorentz transformation $\Lambda$, the momenta $p$ goes to $\Lambda p$, and the spin transforms under the representation $D(W(\Lambda,p))$ of the little group $W$. For massive particles, the little group is the well known group of rotations in three dimensions, $SO(3)$. However, it is also known that $SO(3)$ is homomorphic to $SU(2)$, and the irreducible unitary representations of $SU(2)$ span a Hilbert space of $2j + 1$ dimensions, with $j = n/2$, where $n$ is an integer \cite{Sexl, Tung}. The value of $j$ is what we usually refer to as the spin of the massive particle. In this article, we will be interested in spin-$1/2$ particles, hence the representation of the Wigner rotation is given by \cite{Ahn, Halpern}
\begin{align}
    D(W(\Lambda, p)) & = \frac{(p^0 + m)\cosh(\omega/2) I_{2 \times 2} + (\vec{p} \cdot \hat{e}) \sinh(\omega/2) - i \sinh(\omega/2) \vec{\sigma}\cdot(\vec{p}\times \hat{e}) }{\sqrt{(p^0 + m)((\Lambda p)^0 + m)}}\\
    & =  \cos \frac{\phi}{2} I_{2 \times 2} + i \sin \frac{\phi}{2} (\vec{\sigma} \cdot \hat{n}),
\end{align}
with $I_{2 \times 2}$ being the identity matrix, meanwhile $\vec{\sigma}$ are the Pauli matrices, and
\begin{align}
    & \cos \frac{\phi}{2} = \frac{\cosh(\omega/2)\cosh(\alpha/2) + \sinh(\omega/2)\sinh(\alpha/2) (\hat{e} \cdot \hat{p})}{\sqrt{\frac{1}{2}(1 + \cosh \omega \cosh \alpha + \sinh \omega \sinh \alpha (\hat{e} \cdot \hat{p}))}},\\
    & \sin \frac{\phi}{2} \hat{n} = \frac{ \sinh(\omega/2)\sinh(\alpha/2) (\hat{e} \times \hat{p})}{\sqrt{\frac{1}{2}(1 + \cosh \omega \cosh \alpha + \sinh \omega \sinh \alpha (\hat{e} \cdot \hat{p}))}},
\end{align}
where $\cosh \alpha = p^0/m$, $\omega = \tanh^{-1} v$ is the rapidity of the boost \cite{Rhodes}, $\hat{e}$ is the unit vector pointing in the direction of the boost, $p$ is the $4$-momenta of the particle in $\mathcal{O}$, and $\Lambda p$ is the $4$-momenta of the particle in $\mathcal{O}'$.  As an example, if the momentum is in the $x$-direction of the referece frame $\mathcal{O}$ and the boost is given in the $z$-axis, then 
\begin{equation}
    D(W(\Lambda,p)) = \cos \frac{\phi}{2} I_{2 \times 2} - i \sin \frac{\phi}{2} \sigma_y = \begin{pmatrix}
\cos \frac{\phi}{2}  &  - \sin \frac{\phi}{2} \\
 \sin \frac{\phi}{2} & \cos \frac{\phi}{2}, \\
\end{pmatrix}
\end{equation}
where the Wigner angle $\phi$ is given by
\begin{align}
    & \cos \frac{\phi}{2} = \frac{\cosh(\omega/2)\cosh(\alpha/2)}{\sqrt{\frac{1}{2}(1 + \cosh \omega \cosh \alpha)}},\\
    & \sin \frac{\phi}{2} \hat{n} = \frac{ \sinh(\omega/2)\sinh(\alpha/2) \hat{y}}{\sqrt{\frac{1}{2}(1 + \cosh \omega \cosh \alpha)}}, \\
    & \tan \phi = \frac{\sinh \omega \sinh \alpha}{\cosh \omega + \cosh \alpha},
\end{align}
which implies that $\phi \in [0, \pi/2]$. Hence, the transformation law for the spin-$1/2$ particle with momentum $\vec{p}$ along the $x$-axis of $\mathcal{O}$ is given by
\begin{align}
    & U(\Lambda)\ket{p, 0} = \ket{\Lambda p} \otimes (\cos \frac{\phi}{2} \ket{0} + \sin \frac{\phi}{2} \ket{1}), \\
    & U(\Lambda)\ket{p, 1} = \ket{\Lambda p} \otimes ( -\sin \frac{\phi}{2} \ket{0} + \cos \frac{\phi}{2} \ket{1}),
\end{align}
where $\ket{0}$ means spin `up' and $\ket{1}$ stands for spin `down' along the $z$-axis. Therefore, as one can see, for separable and completely predictable states, a Lorentz boost will only generate superposition between the possible states of the spin of the particle, as already noticed in Ref. \cite{Lanzagorta}.

\section{The Lorentz invariance of complete complementarity relations}
\label{sec:lorccr}

In Ref. \cite{Marcos}, we developed a general framework to obtain complete complementarity relations for a subsystem that belongs to an arbitrary multipartite pure quantum system, just by exploring the purity of the multipartite quantum system.
To make our investigation easier, we begin by assuming that momenta can be treated as discrete variables \cite{Jordan, Friis, Palge}. This can be justified once we can consider narrow distributions centered around different momentum values such
that it is possible to represent them by orthogonal state vectors, i.e., $\braket{p_i}{p_j} = \delta_{i,j} $. Although narrow momenta are an idealization, it is a system worth studying since it helps to understand more realistic systems, and, also, it is possible to approximate continuous momenta as a finite (but large) number of discrete momenta. Also, throughout this article, we will consider only massive particles of spin $1/2$. By doing this, we are considering a particular representation of the Wigner little group. However, the result obtained in this section does not depend on the particular choice of representation, once the representation is unitary.

So, let us consider $n$ massive quantons with spin $1/2$ in a pure state described by $\ket{\Psi}_{A_1,...,A_{2n}} \in \mathcal{H}_{1} \otimes ... \otimes \mathcal{H}_{2n}$ with dimension $d = d_{A_1}d_{A_2}...d_{A_{2n}}$, in the reference frame $\mathcal{O}$. For instance, $A_1$, $A_2$ are referred as the momentum and spin of the first quanton, and so on.  By defining a local orthonormal basis for each degree of freedom (DOF) $A_m$, $\{\ket{i_m}_{A_m}\}_{i = 0}^{d_m - 1}$, $m = 1,...,2n$, the state of the multipartite quantum system can be written as \cite{Mark}
\begin{align}
    \rho_{A_1, ..., A_{2n}} = \ket{\Psi}_{A_1,...,A_{2n}}\bra{\Psi} =  \sum_{i_1,...,i_{2n}} \sum_{j_1,...,j_{2n}} \rho_{i_1 ... i_{2n},j_1...j_{2n}}\ket{i_1,...,i_{2n}}_{A_1,...,A_{2n}}\bra{j_1,...,j_{2n}}.
\end{align}
Without loss of generality, let us consider the state of the DOF $A_1$, which is obtained by tracing over the other subsystems,
\begin{align}
    \rho_{A_1} = \sum_{i_1,j_1}\rho_{i_1,j_1}^{A_1}\ket{i_1}_{A_1}\bra{j_1} = \sum_{i_1,j_1}\sum_{i_2,...,j_{2n}}\rho_{i_1 i_2 ... i_{2n}, j_1 i_2 ... i_{2n}}\ket{i_1}_{A_1}\bra{j_1},
\end{align}
for which the Hilbert-Schmidt quantum coherence and the corresponding predictability measure in terms of the density matrix elements are given by
\begin{align}
    & C_{hs}(\rho_{A_1}) = \sum_{i_1 \neq j_1}\abs{\rho_{i_1,j_1}^{A_1}}^2 = \sum_{i_1 \neq j_1}\abs{\sum_{i_2,...,i_{2n}}\rho_{i_1 i_2 ... i_{2n}, j_1 i_2 ... i_{2n}}}^2, \label{eq:chs} \\
    & P_{l}(\rho_{A_1}) = \sum_{i_1}(\rho_{i_1,i_1}^{A_1})^2 - 1/d_{A_1} = \sum_{i_1}(\sum_{i_2,...,i_{2n}}\rho_{i_1 i_2 ... i_{2n}, i_1 i_2 ... i_{2n}})^2 - 1/d_{A_1}. \label{eq:pl}
\end{align}
Besides, such predictability measure can be first defined as $P_l(\rho_{A_1}) := S^{\max}_l - S_l(\rho_{A_1 diag})$ \cite{Maziero}, where $\rho_{A_1 diag}$ corresponds to the diagonal elements of $\rho_{A_1}$ and $S_l(\rho) := 1 - \Tr \rho^2$ is the linear entropy. To get more intuition about this predictability measure, let us consider the projector onto the state index $i_1$: $\Pi_{i_1} := \ketbra{i_1}$, which can be one of the paths of a Mach-Zehnder interferometer. Now, the uncertainty of the state $i_1$ is given by its variance $\mathcal{V}(\rho_{A_1},\Pi_{i_1}) = \expval{\Pi^2_{i_1}} - \expval{\Pi_{i_1}}^2 =  \rho^{A_1}_{i_1 i_1} - (\rho^{A_1}_{i_1 i_1})^2$ such that sum of the uncertainties of all the possible states (or paths) is given by $\sum_{i_1} \mathcal{V}(\rho_{A_1},\Pi_{i_1}) = 1 - \sum_j (\rho^{A_1}_{i_1 i_1})^2,$ which represents the total uncertainty of all states. One can easily see that $\sum_j \mathcal{V}(\rho_{A_1},\Pi_j)$ is exactly the linear entropy of $\rho_{A_1 diag}$: $S_l(\rho_{A_1 diag}) = 1 - \Tr \rho^2_{A_1 diag}$. Thus, after repeating the same experiment several times, we obtain a probability distribution given by $\rho^{A_1}_{00},..., \rho^{A_1}_{d_{A_1}-1 d_{A_1}-1},$ which represents a probability of the quanton being measured in the state $\ket{0},..., \ket{d_{A_1} - 1}$. From this, it is possible to calculate the uncertainty about the paths through $S_l(\rho_{diag})$ such that $P_{l}(\rho_{A_1 diag}) := S_{l}^{max} - S_{l}(\rho_{A_1 diag})$ offers a measure of the capability to predict what outcome will be obtained in the next run of the experiment. For instance, if  we obtain a uniform  probability distribution, i.e., $\{\rho^{A_1}_{i_1 i_1} = 1 /d_{A_1}\}_{i_1 = 0}^{d_{A_1}-1}$, after repeating the experiment several times, our ability to make a prediction is null. On the other hand, if $\rho_{A_1 diag} \neq I_{A_1}/d_{A_1}$, where $I_{A_1}$ is the identity operator, then $P_{hs}(\rho_{A_1}) \neq 0$. Therefore, it is possible to see that Eq.(\ref{eq:pl})  is a way of quantifying how much the probability distribution expressed by $\rho_{A_1 diag}$ differs from the uniform probability distribution. While the Hilbert-Schmidt quantum coherence can be defined as $C_{hs}(\rho_{A_1}) := \min_{\iota \in \mathcal{I}}||\rho_{A_1}-\iota||_{hs}^{2}$, where $\mathcal{I}$ is the set of all incoherent states (diagonal density operators), and the Hilbert-Schmidt's norm of a matrix $M\in\mathbb{C}^{d \times d}$ is defined as $\norm{M}_{hs}:=  \sqrt{\sum_{j,k} \abs{M_{jk}}^2}$. The minimization procedure yields  Eq.(\ref{eq:chs}). Therefore, the Hilbert-Schmidt quantum coherence is measuring how distant the density operator $\rho_{A_1}$ is in comparison with its closest incoherent state, which in this case is $\rho_{A_1 diag}$, given the Hilbert-Schmidt's norm. Loosely speaking, the quantum coherence is measuring ``how much'' orthogonal superposition is encoded in a given density operator $\rho_{A_1}$ and it is directly related to the off-diagonal elements of $\rho_{A_1}$. Besides, we showed in Ref. \cite{Maziero} that these are bona-fide measures of visibility and predictability, respectively. From these equations, an incomplete complementarity relation, $P_{hs}(\rho_{A_1}) + C_{hs}(\rho_{A_1})  \le (d_{A_1} - 1)/d_{A_1}$, is obtained by exploring the mixedness of $\rho_{A_1}$, i.e., $1 - \Tr \rho_{A_1}^2 \ge 0$. 

Now, since $\rho_{A_1,...,A_{2n}}$ is a pure quantum system, then $1 - \Tr \rho_{A_1,...,A_{2n}}^2 = 0$, or equivalently,
\begin{align}
1 - \Big(\sum_{(i_1,...,i_{2n}) = (j_1,...,j_{2n})} + \sum_{(i_1,...,i_{2n}) \neq (j_1,...,j_{2n})}\Big)\abs{\rho_{i_1 i_2 ... i_{2n}, j_1 j_2 ... j_{2n}}}^2 = 0 \label{eq:pur},    
\end{align}
where
\begin{align}
    \sum_{(i_1,...,i_{2n}) \neq (j_1,...,j_{2n})} = \sum_{\overset{i_1 \neq j_1}{\overset{i_2 = j_2}{\overset{\vdots}{i_{2n} = j_{2n}}}}} + \sum_{\overset{i_1 = j_1}{\overset{i_2 \neq j_2}{\overset{\vdots}{i_{2n} = j_{2n}}}}} + ... + \sum_{\overset{i_1 = j_1}{\overset{i_2 = j_2}{\overset{\vdots}{i_{2n} \neq j_{2n}}}}} + \sum_{\overset{i_1 \neq j_1}{\overset{i_2 \neq j_2}{\overset{\vdots}{i_{2n} = j_{2n}}}}} + ... + \sum_{\overset{i_1 \neq j_1}{\overset{i_2 = j_2}{\overset{\vdots}{i_{2n} \neq j_{2n}}}}} + ... + \sum_{\overset{i_1 \neq j_1}{\overset{i_2 \neq j_2}{\overset{\vdots}{i_{2n} \neq j_{2n}}}}}.
\end{align}
The purity condition (\ref{eq:pur}) can be rewritten as a CCR:
\begin{align}
    P_{l}(\rho_{A_1}) + C_{hs}(\rho_{A_1}) + S_l(\rho_{A_1}) = \frac{d_{A_1} - 1}{d_{A_1}}, \label{eq:ccrhs}
\end{align}
with $S_l(\rho_{A_1})$ being the linear entropy of the subsystem $A_1$ given by
\begin{align}
    S_l(\rho_{A_1}) := \sum_{i_1 \neq j_1} \sum_{(i_2,...,i_{2n}) \neq (j_2,...,j_{2n})}\Big(\abs{\rho_{i_1 i_2 ... i_{2n}, j_1 j_2 ... j_{2n}}}^2 -  \rho_{i_1 i_2 ... i_{2n}, j_1 i_2 ... i_{2n}}\rho_{i_1 j_2 ... j_{2n}, j_1 j_2 ... j_{2n}}^*\Big). \label{eq:linent}
\end{align}
It is worthwhile mentioning the CCR in Eq. (\ref{eq:ccrhs}) is a natural generalization of the complementarity relation obtained by Jakob and Bergou \cite{Jakob, Bergou} for bipartite pure quantum systems. More generally, $E = \sqrt{2S_l(\rho_{A_1})}$, where $E$ is the generalized concurrence obtained in Ref. \cite{Bhaskara} for multi-particle pure states. Now, for the boosted observer $\mathcal{O}$' of Sec. \ref{sec:rep}, the same $n$ massive quantons system is described by $\ket{\Psi_{\Lambda}}_{A_1,...,A_{2n}} = U(\Lambda) \ket{\Psi}_{A_1,...,A_{2n}}$, and the density matrix of the multipartite pure quantum system can be written as \cite{Caban, Vianna}
\begin{equation}
    \rho^{\Lambda}_{A_1, ..., A_{2n}} = \ket{\Psi_{\Lambda}}_{A_1,...,A_{2n}}\bra{\Psi_{\Lambda}} = U(\Lambda) \rho_{A_1, ..., A_{2n}} U^{\dagger}(\Lambda), \label{eq:rho}
\end{equation}
which implies that $\Tr(\rho^{\Lambda}_{A_1, ..., A_{2n}})^2 = \Tr (\rho_{A_1, ..., A_{2n}})^2$, and the whole system remains pure under the Lorentz boost. As we used the purity of the density matrix to obtain the complete complementarity relation, then, from $1 -  \Tr(\rho^{\Lambda}_{A_1, ..., A_{2n}})^2 = 0$, we can obtain
\begin{equation}
    P_{l}(\rho^{\Lambda}_{A_1}) + C_{hs}(\rho^{\Lambda}_{A_1}) + S_l(\rho^{\Lambda}_{A_1}) = \frac{d_{A_1} - 1}{d_{A_1}}.
\end{equation}
This proves our claim that CCR are invariant under Lorentz transformations.  For continuous momenta, the result showed here remains valid if applied to the discrete degrees of freedom with the continuous momenta been traced out, since we defined $P_l, C_{hs}$ and $S_l$ only for the discrete degrees of freedom. Therefore, one can see that there is a transformation law connecting the CCRs for different Lorentz frames.

\section{Relativistic settings}
\label{sec:relset}

\subsection{Single-particle system scenarios}
\label{subsec:single}
\begin{figure}[t]
\subfigure[\footnotesize $S_l(\rho_{\Lambda s}) = S_l(\rho_{\Lambda p})$ as a function of the Wigner angle.]{\includegraphics[width=5.5cm]{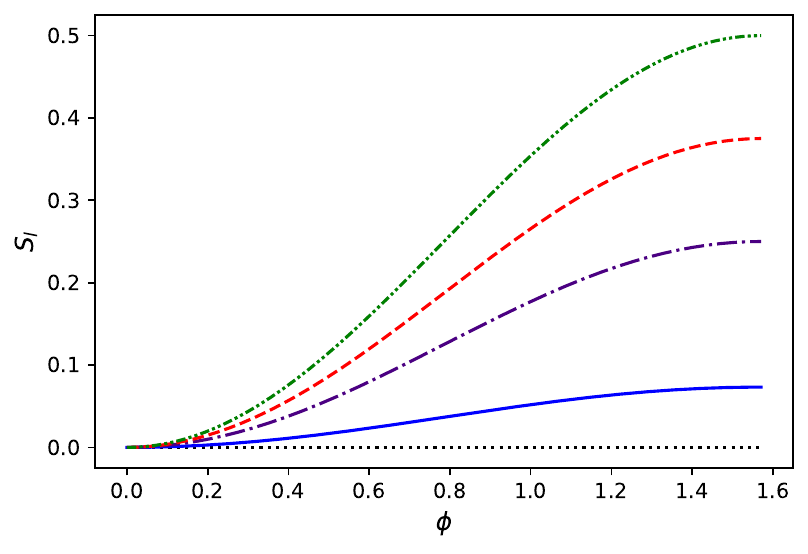}}
\subfigure[\footnotesize  $P_{l}(\rho_{\Lambda s})$ as a function of the Wigner angle.]{\includegraphics[width=5.5cm]{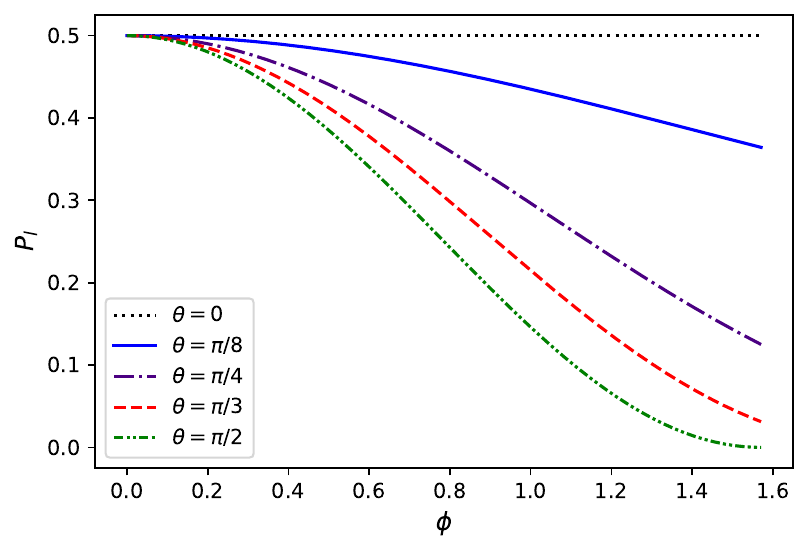}}
\qquad
\subfigure[\footnotesize $C_{hs}(\rho_{\Lambda s})$ as a function of the Wigner angle.]{\includegraphics[width=5.5cm]{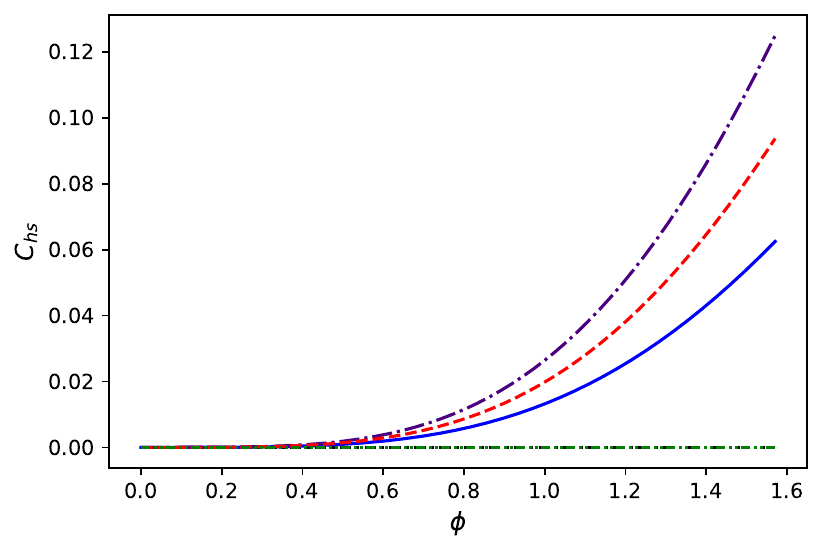}}
\subfigure[\footnotesize  $C_{hs}(\rho_{\Lambda p})$ as a function of the Wigner angle.]{\includegraphics[width=5.5cm]{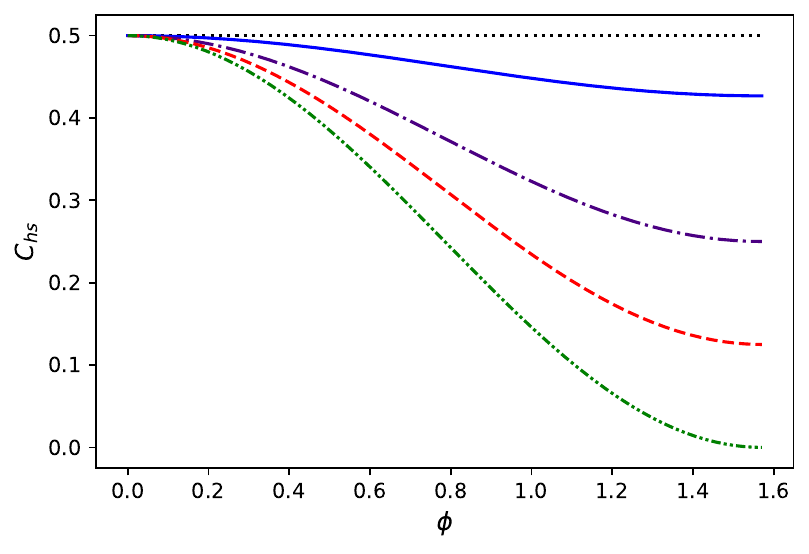}}
\caption{The different aspects of the degrees of freedom in the state $\ket{\Psi_{\Lambda}}$ for different values of $\theta$.}
\label{fig:examp1}
\end{figure}

We begin by considering three different single-particle states where the particle is moving in two opposing directions along the $y$ axis and the spins are aligned with the $z$ axis irrespective of the direction of the boost in the reference frame $\mathcal{O}$:
\begin{align}
    & \ket{\Psi} = \frac{1}{\sqrt{2}}(\ket{p} + \ket{-p}) \otimes \ket{0},  \label{eq:psi}\\
    & \ket{\Xi} = \frac{1}{\sqrt{2}}(\ket{p,0} + \ket{-p,1}), \label{eq:xi}\\
    & \ket{\Phi} = \frac{1}{2}(\ket{p} + \ket{-p}) \otimes ( \ket{0} + \ket{1})\label{eq:phi},
\end{align}
where $\ket{0}$ means spin `up', and $\ket{1}$ spin `down'. In addition, $\ket{-p}$ is describing the state whose spatial momentum has opposite direction in comparison with $\ket{p}$. It is worthwhile mentioning that the states $\ket{\Psi}, \ket{\Phi}$ are separable states, while $\ket{\Xi}$ is a maximal entangled state. Moreover, the state $\ket{\Psi}$ has maximal coherence in the momentum degree of freedom, and maximal predictability in the spin degree of freedom. Meanwhile $\ket{\Phi}$ is maximally coherent in both degrees of freedom, and $\ket{\Xi}$ has no local properties. Now, let us consider an observer $\mathcal{O}$' boosted with velocity $v$ in a direction orthogonal to the momentum of the particle in the frame $\mathcal{O}$, i.e., in the $x-z$ plane, making an angle $\theta \in [0, \pi/2]$ with the $x$-axis. Hence, the direction of boost is given by $\hat{e} = \cos \theta \hat{x} + \sin \theta \hat{z}$, and the Wigner rotation follows directly:
\begin{align}
    D(W(\Lambda, \pm p)) & = \cos \frac{\phi}{2} I_{2 \times 2} + i \sin \frac{\phi}{2} ( \mp \sin \theta \sigma_x \pm \cos \theta \sigma_z)\\
    & = \begin{pmatrix}
\cos \frac{\phi}{2} \pm i \sin \frac{\phi}{2} \cos \theta &  \mp i \sin \frac{\phi}{2} \sin \theta \\
\mp i \sin \frac{\phi}{2} \sin \theta & \cos \frac{\phi}{2} \mp i \sin \frac{\phi}{2} \cos \theta\\
\end{pmatrix}, \label{eq:wigrot}
\end{align}
since $\pm \hat{p} = \pm \hat{y}$. Therefore, the observer in $\mathcal{O}$' assigns in general a different state to the same system. For instance, the state given by Eq. (\ref{eq:psi}) in $\mathcal{O}$' is described by
\begin{align}
    \ket{\Psi_{\Lambda}} & = U(\Lambda) \ket{\Psi} = \frac{1}{\sqrt{2}}( \ket{\Lambda p} \otimes D(W(\Lambda, p)) \ket{0} + \ket{-\Lambda p} \otimes D(W(\Lambda, -p)) \ket{0})\\
    & =  \frac{1}{\sqrt{2}} \Big( \ket{\Lambda p} [(\cos \frac{\phi}{2} + i \sin \frac{\phi}{2}\cos \theta) \ket{0} - i \sin \frac{\phi}{2} \sin \theta \ket{1}] + \ket{-\Lambda p} [(\cos \frac{\phi}{2} - i \sin \frac{\phi}{2} \cos \theta) \ket{0} + i \sin \frac{\phi}{2} \sin \theta \ket{1} ]\Big),
\end{align}
which in general is an entangled state. The reduced density matrix of the spin (momentum) is obtained by tracing out the momentum (spin) states:
\begin{align}
    & \rho_{\Lambda s} = \Tr_{\Lambda p} \ketbra{\Psi_{\Lambda}} =  \begin{pmatrix}
\cos^2 \frac{\phi}{2} + \sin^2 \frac{\phi}{2} \cos^2 \theta &  - \sin^2 \frac{\phi}{2} \sin \theta \cos \theta\\
- \sin^2 \frac{\phi}{2} \sin \theta \cos \theta &  \sin^2 \frac{\phi}{2} \sin^2 \theta\\
\end{pmatrix}, \\
 & \rho_{\Lambda p} = \Tr_{\Lambda s} \ketbra{\Psi_{\Lambda}} = \begin{pmatrix}
\frac{1}{2} &  \frac{1}{2}(\cos \phi + i\sin \phi \cos \theta) \ \\
\frac{1}{2}(\cos \phi - i\sin \phi \cos \theta) & \frac{1}{2}\\
\end{pmatrix}.
\end{align}
In Fig. \ref{fig:examp1}, we plotted the different aspects of the degrees of freedom of the quanton for different values of $\theta$. For instance, if there is no boost, i.e., $\phi = 0$, the state remains unchanged, regardless the direction of the boost. Also, if the boost is along the $x$-axis, $\theta = 0$, the state remains the same. Now, if the boost is along the $z$-axis, the entanglement between the moment and the spin of the particle increases with the increase of the Wigner angle. In exchange, the coherence of the momentum and the predictability of the spin decrease with $\phi$. Beyond that, for any $\theta, \phi \in [0,\pi/2]$, the complete complementarity relation $P_{hs} + C_{hs} + S_{l} = 1/2$ is always satisfied. 

\begin{figure}[t]
\subfigure[\footnotesize $S_l(\rho_{\Lambda j})$, $j = s,p$, as a function of the Wigner angle.]{\includegraphics[width=6cm]{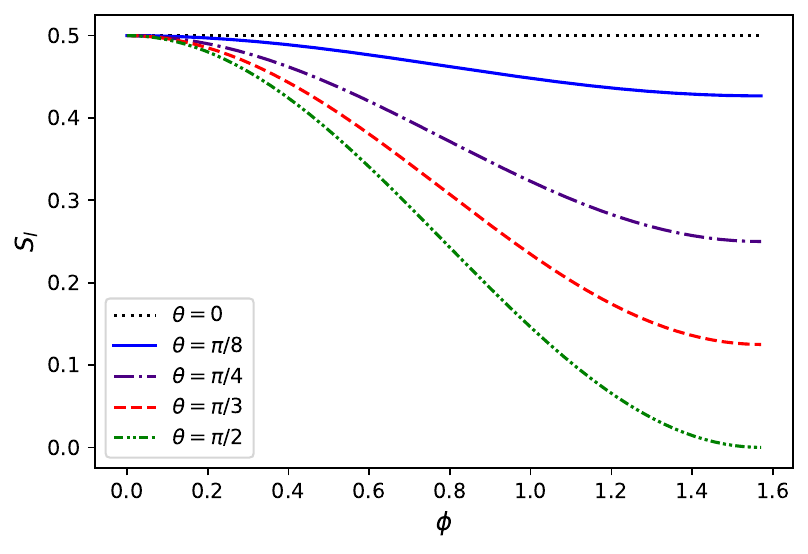}}
\subfigure[\footnotesize  $C_{hs}(\rho_{\Lambda j})$, $j = s,p$, as a function of the Wigner angle.]{\includegraphics[width=6cm]{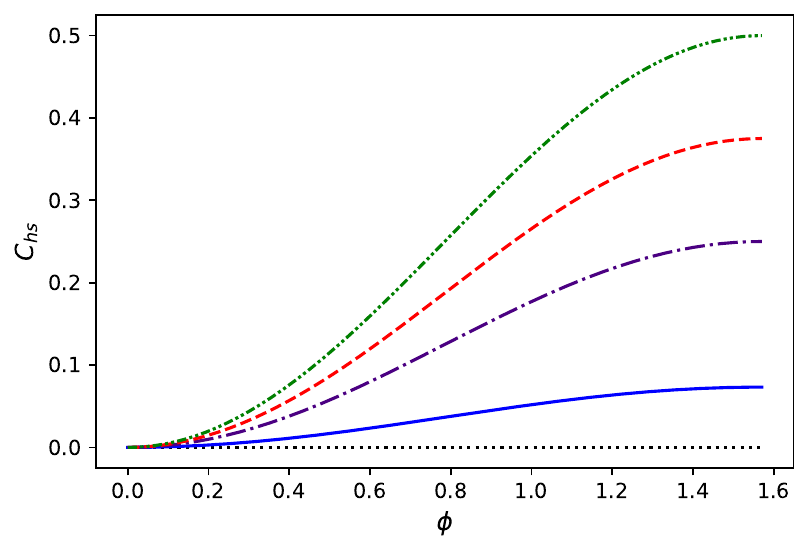}}
\caption{The different aspects of the degrees of freedom of the quanton in state the $\ket{\Xi_{\Lambda}}$ for different values of $\theta$.}
\label{fig:examp2}
\end{figure}

Now, the state $\ket{\Xi}$ given by Eq. (\ref{eq:xi}) is described in $\mathcal{O}'$ as
\begin{align}
    \ket{\Xi_{\Lambda}}  = \frac{1}{\sqrt{2}}\Big( \ket{\Lambda p} [(\cos \frac{\phi}{2} + i \sin \frac{\phi}{2} \cos \theta) \ket{0} - i \sin \frac{\phi}{2}\sin \theta \ket{1}] + \ket{-\Lambda p} [i \sin \frac{\phi}{2}\sin \theta \ket{0} + (\cos \frac{\phi}{2} + i \sin \frac{\phi}{2} \cos \theta) \ket{1}]\Big), 
\end{align}
while the reduced density matrices are given by
\begin{align}
    \rho_{\Lambda s} = \rho^{\dagger}_{\Lambda p} = \begin{pmatrix}
\frac{1}{2} &  i \cos \frac{\phi}{2}\sin \frac{\phi}{2} \sin \theta\ \\
- i \cos \frac{\phi}{2}\sin \frac{\phi}{2} \sin \theta & \frac{1}{2}\\
\end{pmatrix}.
\end{align}
In this example, by inspecting Fig. \ref{fig:examp2}, if there is no boost, i.e., $\phi = 0$ the state remains unchanged, regardless the direction of the boost. Also, if the boost is along the $x$-axis, $\theta = 0$, the state remains the same. However, for $\theta \in (0, \pi/2]$ and $\phi \neq 0$, there is an increase of the coherence of both degrees of freedom, in exchange of the consumption of the entanglement between the momenta and spin of the particle. In the extreme case where $\theta = \pi/2$ and $\phi \to \pi/2$, both degrees of freedom have maximal coherence and the state $\ket{\Xi_{\Lambda}}$ becomes separable
\begin{equation}
    \ket{\Xi_{\Lambda}}_{\phi = \theta = \pi/2} = \frac{1}{2} ( \ket{\Lambda p} + i\ket{- \Lambda p}) \otimes ( \ket{0} - i\ket{1}).
\end{equation}

\begin{figure}[t]
\subfigure[\footnotesize $S_l(\rho_{\Lambda s})$ as a function of the Wigner angle.]{\includegraphics[width=5.5cm]{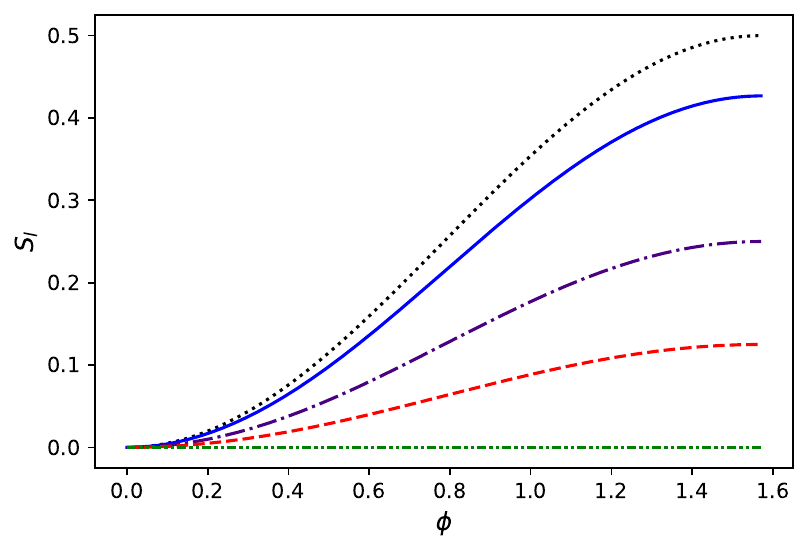}}
\subfigure[\footnotesize $P_l(\rho_{\Lambda s})$ as a function of the Wigner angle.]{\includegraphics[width=5.5cm]{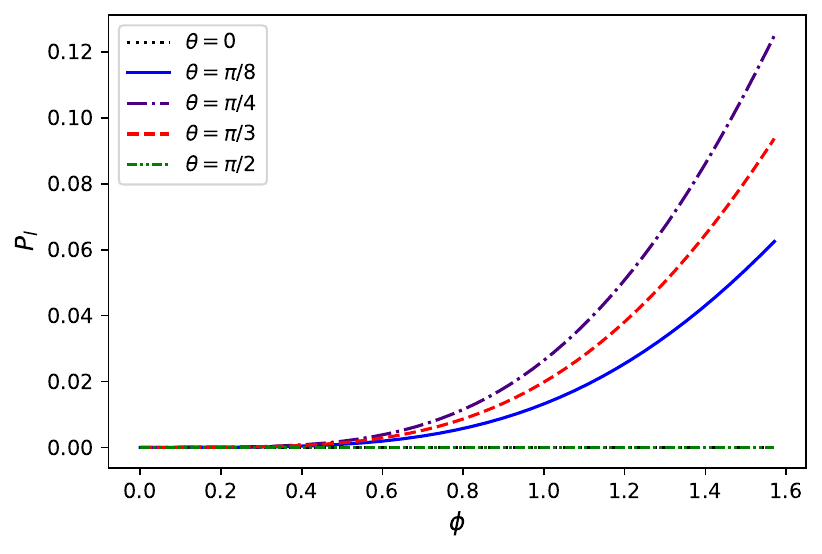}}
\subfigure[\footnotesize  $C_{hs}(\rho_{\Lambda s})$ as a function of the Wigner angle.]{\includegraphics[width=5.5cm]{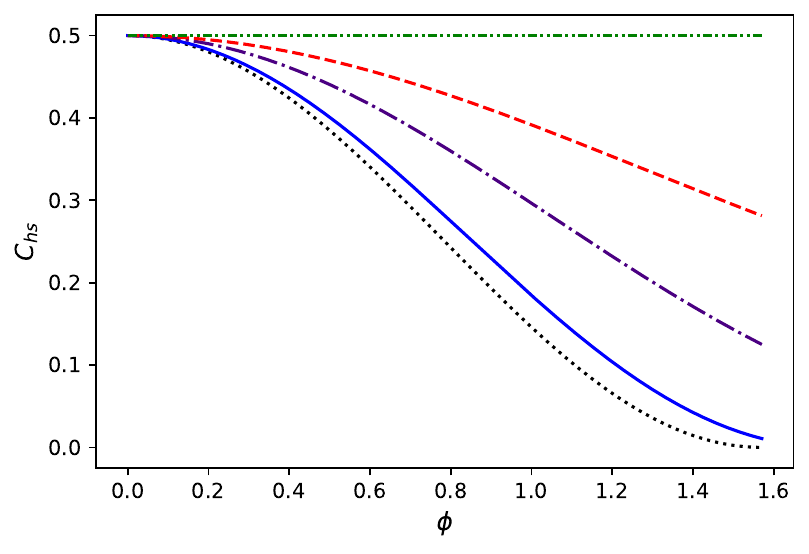}}
\caption{The different aspects of the spin of the quanton in the state $\ket{\Phi_{\Lambda}}$ for different values of $\theta$.}
\label{fig:examp3}
\end{figure}

Lastly, in the boosted frame $\mathcal{O}'$, the state $\ket{\Phi}$ given by Eq. (\ref{eq:phi}) is described by
\begin{align}
    \ket{\Phi_{\Lambda}} & = \frac{1}{2}\Big( \ket{\Lambda p}\{[\cos \frac{\phi}{2} + i \sin \frac{\phi}{2}( \cos \theta - \sin \theta)]\ket{0} + [\cos \frac{\phi}{2} - i \sin \frac{\phi}{2}( \cos \theta + \sin \theta)]\ket{1} \} \\ 
    & \hspace{0.9cm} + \ket{- \Lambda p} \{[\cos \frac{\phi}{2} - i \sin \frac{\phi}{2}( \cos \theta - \sin \theta)]\ket{0} + [\cos \frac{\phi}{2} + i \sin \frac{\phi}{2}( \cos \theta + \sin \theta)]\ket{1} \}\Big),
\end{align}
with the reduced density matrices being
\begin{align}
    & \rho_{\Lambda s}  = \begin{pmatrix}
\frac{1}{2} & \frac{1}{2}(\cos^2 \frac{\phi}{2} - \sin^2 \frac{\phi}{2} \cos 2\theta)  \ \\
\frac{1}{2}(\cos^2 \frac{\phi}{2} - \sin^2 \frac{\phi}{2} \cos 2\theta) & \frac{1}{2}\\
\end{pmatrix}, \\
& \rho_{\Lambda p}  = \begin{pmatrix}
\frac{1}{2} & \frac{1}{2}(\cos \phi - i \sin \phi \sin \theta)  \ \\
\frac{1}{2}(\cos \phi + i \sin \phi \sin \theta) & \frac{1}{2}\\
\end{pmatrix}.
\end{align}
In contrast with the second example, here the entanglement between momentum and spin increases with the Wigner angle, in exchange of the consumption of the coherence of both degrees of freedom. However, for $\theta = \pi/2$ the state is separable:
\begin{equation}
    \ket{\Phi_{\Lambda}}_{\phi = \pi/2} = \frac{1}{2}( A \ket{\Lambda p} + A^* \ket{-\Lambda p}) \otimes (\ket{0} + \ket{1}),
\end{equation}
where $A = cos \frac{\phi}{2} - i \sin \frac{\phi}{2}$ and $A^{*}$ is the complex conjugate of $A$. The coherence and entropy of the momentum has the same qualitative behavior as for the spin, which is plotted in Fig. \ref{fig:examp3}. 

From these examples, we can see if the state of the system is separable, and has no superposition between the momentum states in the reference frame $\mathcal{O}$, then a Lorentz boost cannot generate entanglement between the momentum and spin degrees of freedom. This result helps to explain the reported generation of entanglement between momenta and spin in one of the first studies of single-particle systems carried out by Peres et al. \cite{Peres}, where they considered a particle with mass $m$ whose momentum wave function in the rest frame is given by $\psi(\vec{p}) =(2\pi)^{-3/4}w^{3/2} e^{-\vec{p}^2/2w^2}$, with $w$ being the width of the wave packet. This is a Gaussian state of minimum uncertainty, but still represents a continuous superposition. Therefore, in this case it is possible to generate entanglement via a Lorentz transformation.

\subsection{Two-particle system scenarios}
\label{subsec:two}

As showed in Sec. \ref{subsec:single}, if the momentum states have no coherence, then it is not possible to generate entanglement between the momenta and spin degrees of freedom. In this section, we begin by discussing this issue for the two-particle case and end this section giving two examples. Now, let us consider a two-particle state described in $\mathcal{O}$ as
\begin{equation}
    \ket{\Psi}_{A,B} = \sum_{\sigma, \lambda} \psi_{\sigma} \psi_{\lambda} \ket{p,q}_{A,B} \otimes \ket{\sigma, \lambda}_{A,B},
\end{equation}
where $\sum_{j} \abs{\psi_{j}}^2 = 1$ for $j = \sigma, \lambda$. In addition, $\ket{p,q}_{A,B} = \ket{p}_A \otimes \ket{q}_B$ denotes the momentum state of particle A and B, respectively, meanwhile $ \ket{\sigma, \lambda}_{A,B} = \ket{\sigma}_A \otimes \ket{\lambda}_B$ represents the state of the spins of the particle A and B. The state $ \ket{\Psi}_{A,B}$ is separable and has no coherence between momentum states in the reference frame $\mathcal{O}$. Now, in the boosted frame $\mathcal{O}'$, we have $\ket{\Psi_{\Lambda}}_{A,B} = U(\Lambda) \ket{\Psi}_{A,B}$, i.e.,
\begin{align}
    \ket{\Psi_{\Lambda}}_{A,B} & = \sum_{\sigma, \lambda} \psi_{\sigma} \psi_{\lambda} \ket{\Lambda p, \Lambda q}_{A,B} \otimes D(W(\Lambda, p))\ket{\sigma}_A \otimes D(W(\Lambda, q))\ket{\lambda}_B\\
    & = \ket{\Lambda p, \Lambda q}_{A,B} \otimes \sum_{\sigma} \psi_{\sigma} D(W(\Lambda, p))\ket{\sigma}_A \otimes \sum_{\lambda} \psi_{\lambda}D(W(\Lambda, q))\ket{\lambda}_B,
\end{align}
which is also separable. In this case, the Wigner rotation will only change the coherences of the spin states of the particles A and B. Now, let's consider a state in $\mathcal{O}$ with superposition in the momentum states of the particle A
\begin{equation}
    \ket{\Phi}_{A,B} = \sum_{p} \psi(p) \ket{p,q}_{A,B} \otimes \ket{\sigma, \lambda}_{A,B},
\end{equation}
with $\sum_p \abs{ \psi(p)}^2 = 1$. Then, a Lorentz boost can generate entanglement between the momentum and spin of the particle A
\begin{align}
    \ket{\Phi_{\Lambda}}_{A,B} = \sum_{p} \psi_{\sigma}(p) \ket{\Lambda p}_A \otimes D(W(\Lambda,p)) \ket{\sigma}_A \otimes \ket{\Lambda q}_B \otimes D(W(\Lambda, q)) \ket{\lambda}_B.
\end{align}
However, there is no entanglement between particles A and B. Similarly, if we consider that the state in $\mathcal{O}$ has coherence in the momentum states of A and B, there will be no entanglement between particles A and B. To obtain an entangled state of the whole system in $\mathcal{O}'$, we have to consider a state in $\mathcal{O}$ already entangled in the momentum degrees of freedom, i.e.,
\begin{equation}
    \ket{\Xi}_{A,B} = \sum_{p,q} \psi(p,q) \ket{p,q}_{A,B} \otimes \ket{\sigma, \lambda}_{A,B},
\end{equation}
with $\sum_{p,q} \abs{\psi(p,q)}^2 = 1$. Hence, in boosted frame $\mathcal{O}'$ we have
\begin{align}
    \ket{\Xi_{\Lambda}}_{A,B} = \sum_{p,q} \psi(p,q) \ket{\Lambda p, \Lambda q}_{A,B} \otimes D(W(\Lambda, p))\ket{\sigma}_A \otimes D(W(\Lambda, q))\ket{\lambda}_B. 
\end{align}

\begin{figure}[t]
\subfigure[\footnotesize $S_l(\rho_{\Lambda s})$ as a function of the Wigner angle.]{\includegraphics[width=5.5cm]{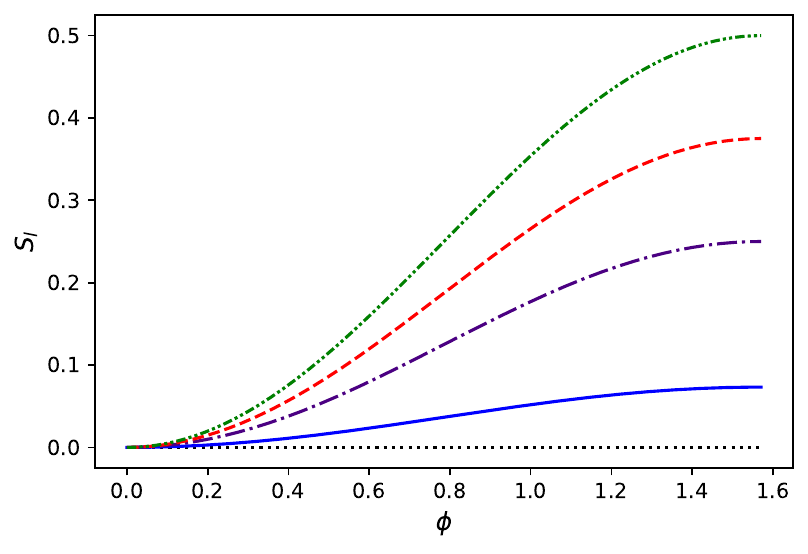}{\label{fig:4a}}}
\subfigure[\footnotesize $P_l(\rho_{\Lambda s})$ as a function of the Wigner angle.]{\includegraphics[width=5.5cm]{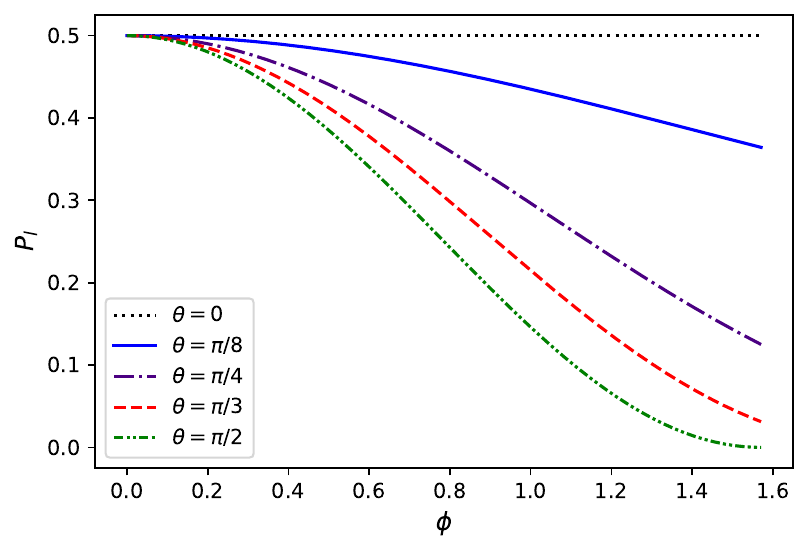}{\label{fig:4b}}}
\subfigure[\footnotesize  $C_{hs}(\rho_{\Lambda s})$ as a function of the Wigner angle.]{\includegraphics[width=5.5cm]{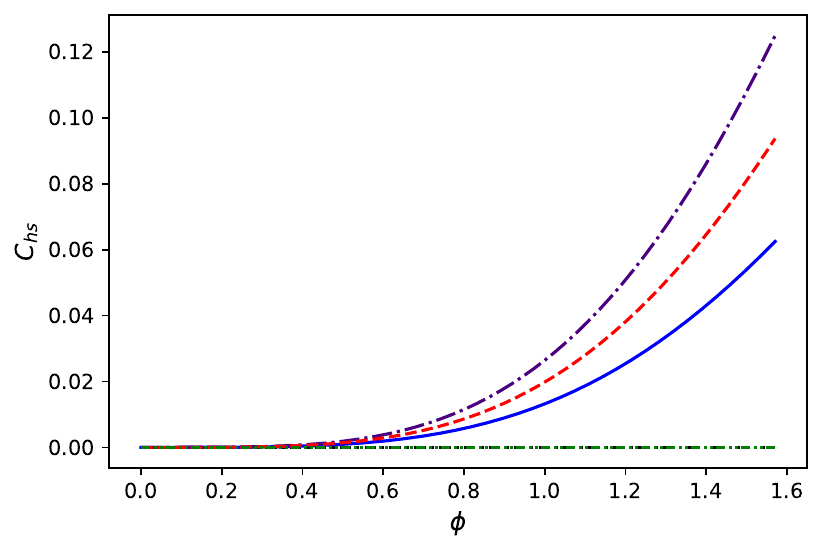}{\label{fig:4c}}}
\subfigure[\footnotesize $C_{hs}(\rho_{\Lambda p \Lambda p})$ as a function of the Wigner angle.]{\includegraphics[width=5.5cm]{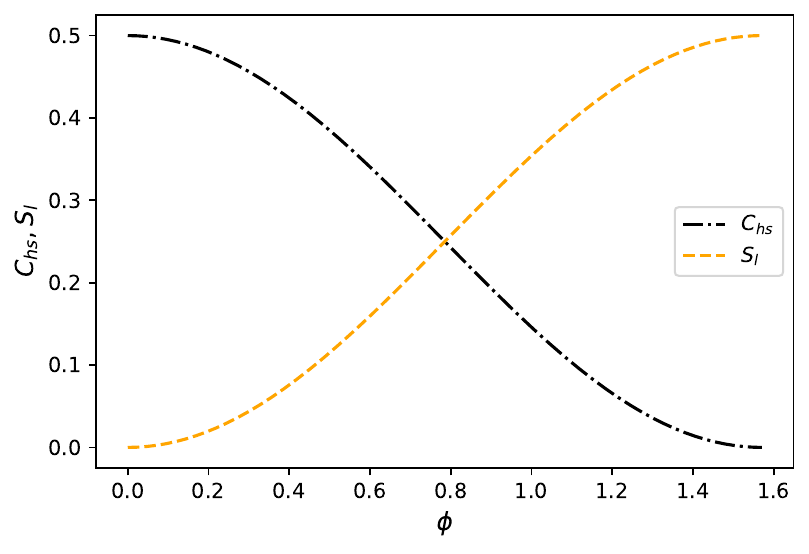}{\label{fig:4d}}}

\caption{The different aspects of $\ket{\Xi_{\Lambda}}$ for different values of $\theta$.}
\label{fig:examp4}
\end{figure}

For instance, if we consider the particles moving in two opposing directions along the y axis and the spins are aligned with the $z$ axis irrespective of the direction of the boost in the reference frame $\mathcal{O}$,
\begin{equation}
    \ket{\Xi}_{A,B} = \frac{1}{\sqrt{2}}(\ket{p, -p}+ \ket{-p, p}) \otimes \ket{0,0},
\end{equation}
as before, the observer $\mathcal{O}$' is boosted with velocity $v$ in a direction orthogonal to the momentum of the particle in the frame $\mathcal{O}$, i.e., in the $x-z$ plane, making an angle $\theta \in [0, \pi/2]$ with the $x$-axis. Then, the Wigner rotation is given by Eq. (\ref{eq:wigrot}) and the state in the boosted frame is described by
\begin{align}
     \ket{\Xi_\Lambda}_{A,B} & = \frac{1}{\sqrt{2}}(\ket{\Lambda p, - \Lambda p} + \ket{-\Lambda p, \Lambda p}) \otimes [(\cos^2 \frac{\phi}{2} + \sin^2 \frac{\phi}{2} \cos^2 \theta) \ket{0,0} + \sin^2 \frac{\phi}{2}\sin^2 \theta \ket{1,1} \nonumber \\ 
     & \hspace{0.4cm} - \sin^2 \frac{\phi}{2} \sin \theta \cos \theta (\ket{0,1} + \ket{1,0})] + \frac{i \cos \frac{\phi}{2} \sin \frac{\phi}{2} \sin \theta}{\sqrt{2}}(\ket{\Lambda p, - \Lambda p} - \ket{-\Lambda p, \Lambda p}) \otimes (\ket{0,1} - \ket{1,0}),
\end{align}
which is an entangled state between all degrees of freedom. In this case, the resource consumed to generate entanglement of the spins of the particles is the bipartite coherence of the reduced momentum-momentum density matrix of the particles A and B
\begin{align}
    \rho_{\Lambda p, \Lambda p} = \frac{1}{2}(\ketbra{\Lambda p, - \Lambda p} + \ketbra{- \Lambda p, \Lambda p}) + \frac{1}{2}(\cos^4 \frac{\phi}{2} + \sin^4 \frac{\phi}{2})(\ketbra{\Lambda p, - \Lambda p}{- \Lambda p, \Lambda p} + t.c.),
\end{align}
where t.c. stands of transpose conjugated. In Fig. \ref{fig:4d}, we plotted the coherence and the linear entropy of $\rho_{\Lambda p, \Lambda p}$ as a function of $\phi$, where $S_l(\rho_{\Lambda p, \Lambda p})$ is measuring the entanglement of $\rho_{\Lambda p, \Lambda p}$ as a whole with rest of the degrees of freedom. Meanwhile, the concurrence measure \cite{Woott} $E$ of $\rho_{\Lambda p, \Lambda p}$ decreases monotonically with Wigner angle, which means the momentum-momentum entanglement decreases with $\phi$, once $E(\rho_{\Lambda p \Lambda p}) = \sqrt{2 C_{hs}(\rho_{\Lambda p \Lambda p})}$ . In addition, Figs. \ref{fig:4a}, \ref{fig:4b} and \ref{fig:4c} represent the behavior of the different aspects of the spin of particle A. The aspects of the spin of particle B display similar behavior. It is worth emphasizing that it is not the entanglement between spin-spin that increases, but the entanglement of the spin of one of the particles with all the other degrees of freedom. In Ref. \cite{Adami}, the authors discussed the generation of spin-spin entanglement for two particles under Lorentz boosts.  Meanwhile, the momentum state of particles A and B are given by $\rho^A_{\Lambda p} = \rho^B_{\Lambda p} = \frac{1}{2}(\ketbra{\Lambda p} + \ketbra{- \Lambda p})$, which implies that $S_l(\rho^j_{\Lambda p}) = 1/2$, $j = A, B$, and the overall entanglement between the momentum of particle A (B) with rest of the system does not change under the Lorentz boost, even though the entanglement of momentum-momentum decreases. Hence, in this case, the entanglement of the momentum of particle A (B) is redistributed among the others degrees of freedom. For $\phi = 0$ (no boost), just the momentum of the particles are entangled. In the limit $\phi = \pi/2$, the momentum of the particles are entangled with the spins, therefore the entanglement of momentum-momentum has to decrease.

\begin{figure}[t]
\subfigure[\footnotesize $C_{hs}(\rho_{\Lambda s}), S_{l}(\rho_{\Lambda s})$ as a function of the Wigner angle.]{\includegraphics[width=6cm]{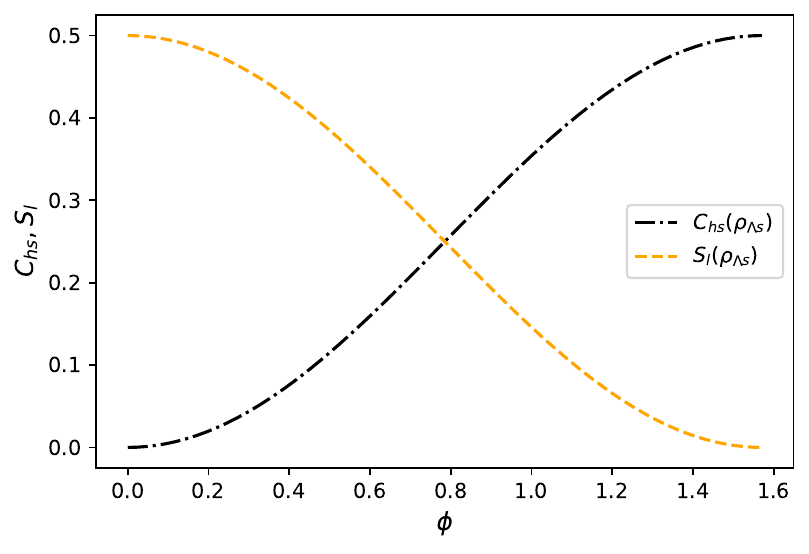}{\label{fig:last1}}}
\subfigure[$C_{hs}(\rho_{\Lambda p \Lambda p}), S_{l}(\rho_{\Lambda p \Lambda p})$ as a function of the Wigner angle.]{\includegraphics[width=6cm]{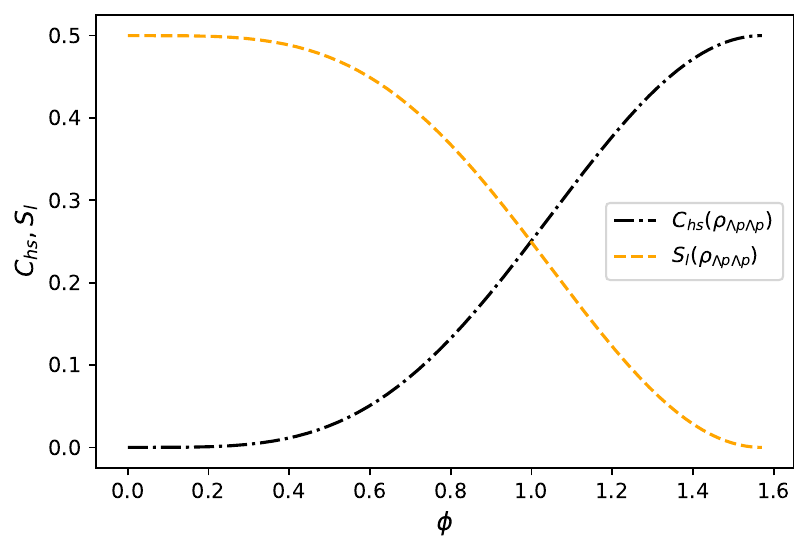}{\label{fig:last2}}}
\caption{The different aspects of $\ket{\Upsilon_{\Lambda}}$.}
\label{fig:last}
\end{figure}

In contrast, we also can redistribute entanglement to generate coherence in the spin states of the particles A and B. For instance, let us consider the following two-particle states in $\mathcal{O}$
\begin{align}
    & \ket{\Upsilon}_{A,B} =  \frac{1}{\sqrt{2}}(\ket{p,0}_{A} \otimes \ket{-p,1}_B + \ket{-p,1}_{A} \otimes \ket{p,0}_B) = \frac{1}{\sqrt{2}}(\ket{p,-p}_{A,B} \otimes \ket{0,1}_{A,B} + \ket{-p,p}_{A,B} \otimes \ket{1,0}_{A,B}),
\end{align}
with the momentum of the particles along the $y$-axis. Now, for a boosted frame $\mathcal{O}$' along the $z$-axis, the Wigner rotation is given by Eq. (\ref{eq:wigrot}) imposing $\theta = \pi/2$. Hence,
\begin{align}
      \ket{\Upsilon_{\Lambda}}_{A,B} & =  \frac{1}{\sqrt{2}}i \cos \frac{\phi}{2} \sin \frac{\phi}{2}(\ket{\Lambda p,- \Lambda p} + \ket{- \Lambda p, \Lambda p}) \otimes( \ket{0,0} - \ket{1,1}) + \frac{1}{\sqrt{2}}\ket{\Lambda p,- \Lambda p}\otimes(\cos^2 \frac{\phi}{2} \ket{0,1} \nonumber \\ & + \sin^2 \frac{\phi}{2} \ket{1,0}) +  \frac{1}{\sqrt{2}}\ket{- \Lambda p, \Lambda p}\otimes(\sin^2 \frac{\phi}{2} \ket{0,1} + \cos^2 \frac{\phi}{2} \ket{1,0}). 
\end{align}
The reduced spin density matrices of each particle are given by
\begin{align}
    & \rho^A_{\Lambda s} = \rho^B_{\Lambda s} = \begin{pmatrix}
\frac{1}{2} & i\cos \frac{\phi}{2} \sin \frac{\phi}{2}  \ \\
-i\cos \frac{\phi}{2} \sin \frac{\phi}{2} & \frac{1}{2}
\end{pmatrix}, 
\end{align}
and $\rho^A_{\Lambda p} = \rho^B_{\Lambda p} = \frac{1}{2} I_{2 \times 2}$, where $I_{2 \times 2}$ is the identity matrix. The entanglement of the spin of the particle A (B) with the rest of the system decreases with the Wigner angle. In exchange, the coherence of the spin of particle A (B) increases, as shown in Fig. \ref{fig:last1}. In addition, the entanglement of $\rho_{\Lambda p \Lambda p}$ as a whole with the spins of the particles also decreases with $\phi$. From Fig. \ref{fig:last2}, the bipartite coherence increases, since it is related to the momentum-momentum entanglement of particle A and B, once $E(\rho_{\Lambda p \Lambda p}) = \sqrt{2 C_{hs}(\rho_{\Lambda p \Lambda p})}$, as we also can see from 
\begin{align}
    \rho_{\Lambda p \Lambda p} = \frac{1}{2}\left(\ketbra{\Lambda p, - \Lambda p} + \ketbra{- \Lambda p, \Lambda p}\right) + \left(2 \cos^2 \frac{\phi}{2}\sin^2 \frac{\phi}{2} \ketbra{\Lambda p, - \Lambda p}{-\Lambda p, \Lambda p} + t.c.\right). 
\end{align}

Hence, the entanglement of the momentum of the particle A (B) with rest of the degrees of the system remains the same under the Lorentz boost, although it is shuffled around among the degrees of freedom. For instance, when $\phi = 0$ (no boost), the momentum of particle A is entangled with all the others degrees of freedom. However, in the limit $\phi = \pi/2$, the momentum of the particle A is entangled just with the momentum of the particle B, since
\begin{align}
    \ket{\Upsilon_{\Lambda_{\phi = \pi/2}}}_{A,B} = \frac{1}{\sqrt{2}}(\ket{\Lambda p, - \Lambda p}_{A,B} + \ket{- \Lambda p, \Lambda p}_{A,B}) \otimes \frac{1}{\sqrt{2}} (i\ket{0}_A + \ket{1}_A) \otimes \frac{1}{\sqrt{2}}(\ket{0}_B - i \ket{1}_B),
\end{align}
and $S_l(\rho_{\Lambda p \Lambda p}) = 0$ for $\phi = \pi/2$.

\section{Conclusions}
\label{sec:con}
 Although the entanglement entropy does not remain invariant under Lorentz boosts, and neither do the measures of predictability and coherence, we showed in this work that these three measures taken together, in a complete complementarity relation (CCR), are Lorentz invariant. Even though it is possible to formally define spin in any Lorentz frame, there is no relationship between the observable expectation values in different Lorentz frames, according to Peres et. al. \cite{Peres}. Here the situation is quite different. First, it is possible to formally define complementarity in any Lorentz frame and, in principle, there is no relationship between the complementarity relations in different Lorentz frames. However, our results showed that it is possible indeed to connect complete complementarity relations in different Lorentz frames. Therefore, we showed how the connection between the CCR defined in different Lorentzian frames is possible, and disclosed interesting aspects of the redistribution of quantum features for the relativistic dynamics of one- and two-particle states and how the role of CCRs helps to keep tracking of how this redistribution of quantum features is done.

\begin{acknowledgments}
This work was supported by the Coordena\c{c}\~ao de Aperfei\c{c}oamento de Pessoal de N\'ivel Superior (CAPES), process 88882.427924/2019-01, and by the Instituto Nacional de Ci\^encia e Tecnologia de Informa\c{c}\~ao Qu\^antica (INCT-IQ), process 465469/2014-0.
\end{acknowledgments}

\end{document}